\documentclass[letter]{aa} 

\usepackage{graphicx}
\usepackage{txfonts}

\usepackage{natbib}
\usepackage{aas_macros}
\usepackage{color}
\usepackage{bm}

\newcommand{\tred}[1]{#1}

\newcommand{\figdir}{./}
\newcommand{\figprd}{\figdir/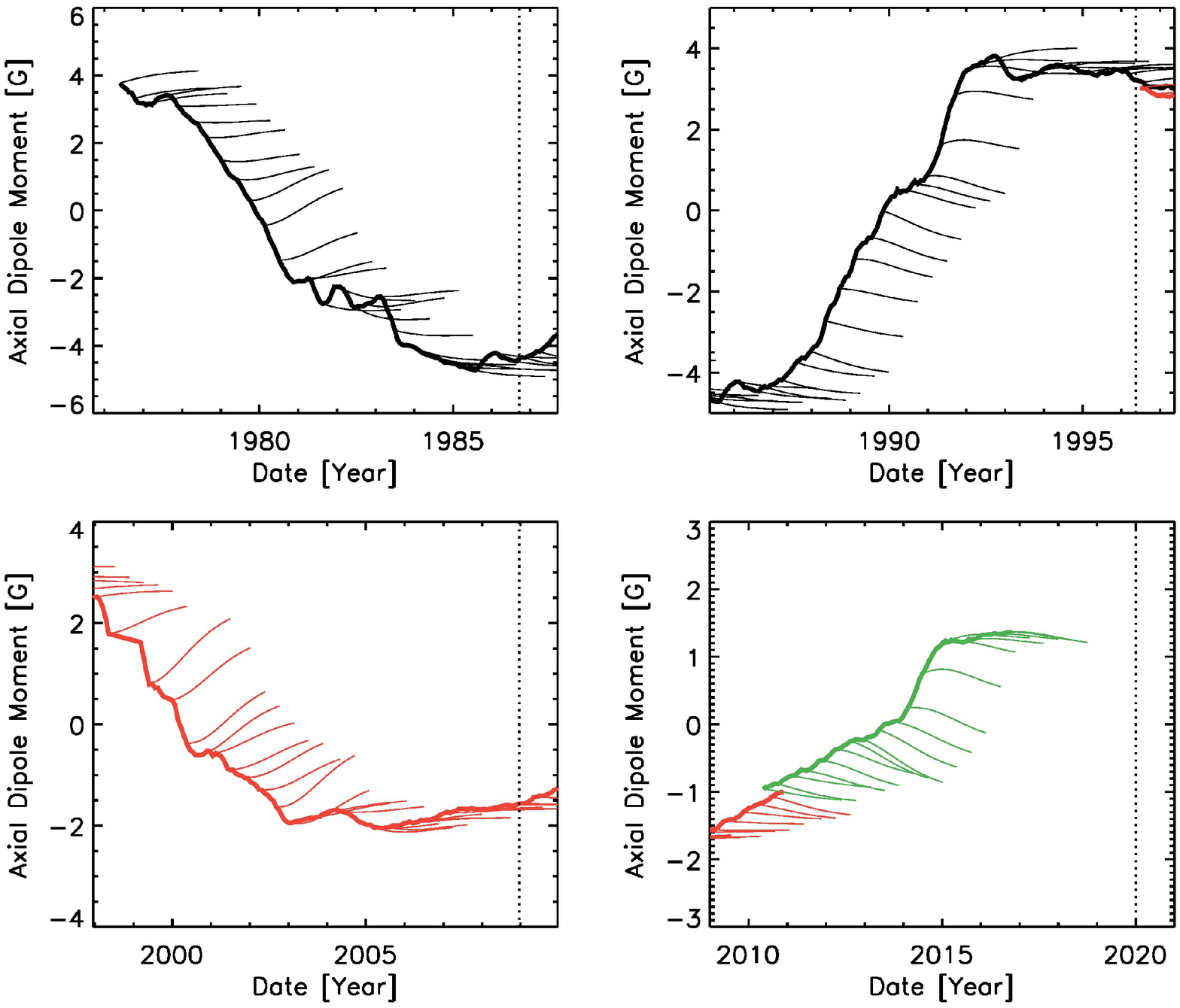}
\newcommand{\figssn}{\figdir/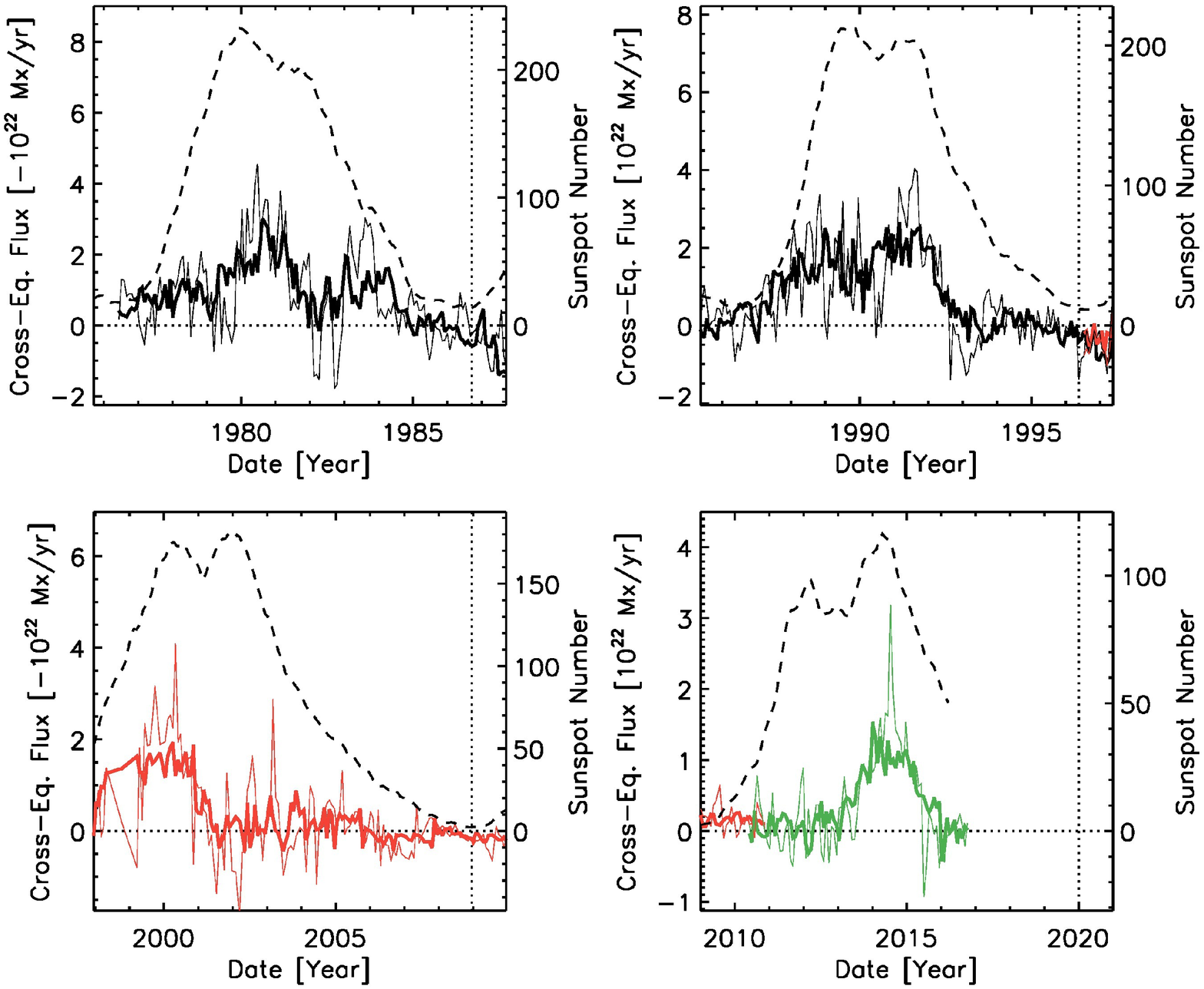}
\newcommand{\figcycle}{\figdir/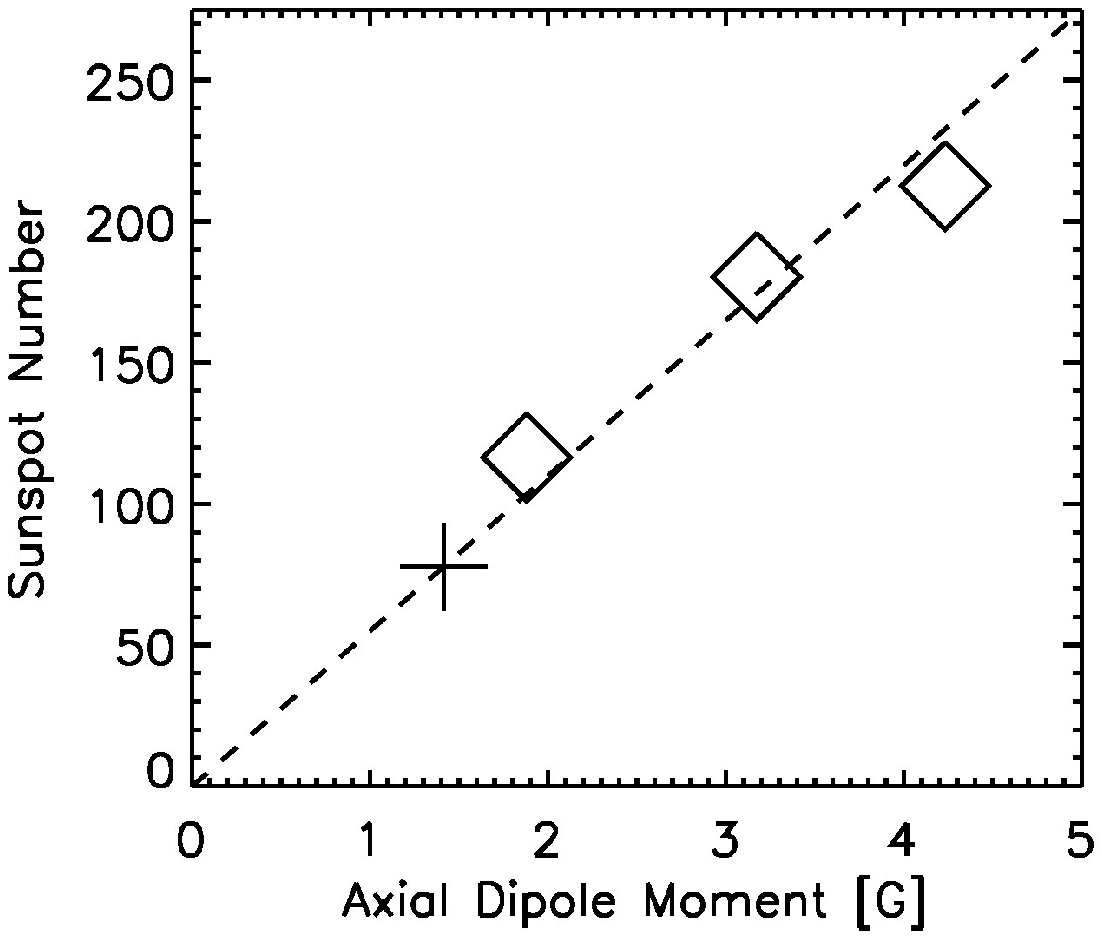}

\begin{document}

\title{
Improvement of Solar Cycle Prediction:\\
Plateau of Solar Axial Dipole Moment
}

\titlerunning{Plateau of Solar Axial Dipole Moment}




\author{
H. Iijima\inst{1}
\and
H. Hotta\inst{2}
\and
S. Imada\inst{1}
\and
K. Kusano\inst{1}
\and
D. Shiota\inst{3}
}

\institute{
Division for Integrated Studies,
Institute for Space-Earth Environmental Research, Nagoya University,
Furocho, Chikusa-ku, Nagoya, Aichi 464-8601, Japan\\
\email{h.iijima@isee.nagoya-u.ac.jp}
\and
Department of Physics, Faculty of Science, Chiba University,
1-33 Yayoi-chou, Inage-ku, Chiba 263-8522, Japan
\and
Applied Electromagnetic Research Institute,
National Institute of Information and Communications Technology (NICT),
4-2-1 Nukui-Kitamachi, Koganei, Tokyo 184-8795, Japan
}

\date{Received; accepted}

\abstract
{}
{
We report the small temporal variation
of the axial dipole moment near the solar minimum
and its application to the solar cycle prediction
by the surface flux transport (SFT) model.
}
{
We measure the axial dipole moment
using the photospheric synoptic magnetogram
observed by the Wilcox Solar Observatory (WSO),
the ESA/NASA Solar and Heliospheric Observatory
Michelson Doppler Imager (MDI),
and the NASA Solar Dynamics Observatory
Helioseismic and Magnetic Imager (HMI).
We also use the surface flux transport model
for the interpretation and prediction
of the observed axial dipole moment.
}
{
We find that the observed axial dipole moment
becomes approximately constant
during the period of
several years before each cycle minimum,
which we call the axial dipole moment plateau.
The cross-equatorial magnetic flux transport
is found to be small during the period,
although the significant number of sunspots
are still emerging.
The results indicates that the newly emerged
magnetic flux does not contributes to the build up
of the axial dipole moment near the end of each cycle.
This is confirmed by showing that
the time variation of the observed axial dipole moment
agrees well with that predicted by the SFT model
without introducing new emergence of magnetic flux.
These results allows us to predict the axial dipole moment
in Cycle 24/25 minimum using the SFT model
without introducing new flux emergence.
The predicted axial dipole moment of Cycle 24/25 minimum
is 60--80 percent of Cycle 23/24 minimum,
which suggests the amplitude of Cycle 25
even weaker than the current Cycle 24.
}
{
The plateau of the solar axial dipole moment is an important feature
for the longer prediction of the solar cycle based on the SFT model.
}


\maketitle

\section{Introduction}

The prediction of the 11-year sunspot cycle
is an important task for the long-term prediction of the space weather.
Various methods have been suggested
on this subject \citep{1999JGR...10422375H,2010LRSP....7....6P,
2012SoPh..281..507P,2016SpWea..14...10P}.
However, the predictability of these methods is still controversial,
especially on the prediction of weak sunspot activity in Cycle 24.
One of the most successful approaches is the precursor method
\citep[e.g.,][]{ohl1966wolf,1978GeoRL...5..411S,
2005GeoRL..32.1104S,2005GeoRL..3221106S,
2009ApJ...694L..11W,2013PhRvL.111d1106M}.
This method uses
the strength of the polar magnetic field on the solar surface
(or relating indices like the axial dipole moment and
geomagnetic activity index)
in the solar cycle minimum as an indicator.
This indicator has a high correlation
with the amplitude of the next cycle maximum.
The precursor method allows us
to predict the next solar cycle amplitude
approximately five years before the cycle maximum.

Recently, the prediction of the polar field in Cycle 24/25 minimum
is carried out based on the surface flux transport (SFT) models
\citep{2014ApJ...780....5U,2016ApJ...823L..22C,2016arXiv161105106H}.
They assume empirical modelings
to predict the newly emerging magnetic flux
until the solar cycle minimum.
As noted by \cite{2014ApJ...791....5J,2015ApJ...808L..28J},
modeling of the new flux emergence
causes the considerable uncertainty (or dispersion)
about the resulting axial dipole moment.
The accuracy and validity of the emerging flux modeling
is crucial for the prediction of the future solar cycle.
In this study,
we investigate the time evolution of the observed
axial dipole moment
to explore the better modeling of the new flux emergence
in the SFT model.
We also carry out the prediction of the next solar cycle amplitude
based on the observation and the simulation by the SFT model.

\section{Observation\label{sec:observation}}

We use the line-of-sight synoptic magnetogram
taken from the ESA/NASA Solar and Heliospheric Observatory
Michelson Doppler Imager \citep[MDI]{1995SoPh..162..129S},
the NASA Solar Dynamics Observatory
Helioseismic and Magnetic Imager \citep[HMI]{2012SoPh..275..207S},
and the Wilcox Solar Observatory (WSO).
The line-of-sight magnetic field is projected
to the radial component of the magnetic field
assuming that the photospheric magnetic field is vertical.
The magnetic field strength of the WSO data
is multiplied by a factor of two
for accounting the saturation of the magnetic signal.
We also use the monthly sunspot number provided by WDC-SILSO.

\begin{figure}[!t]
 \includegraphics[width=\hsize]{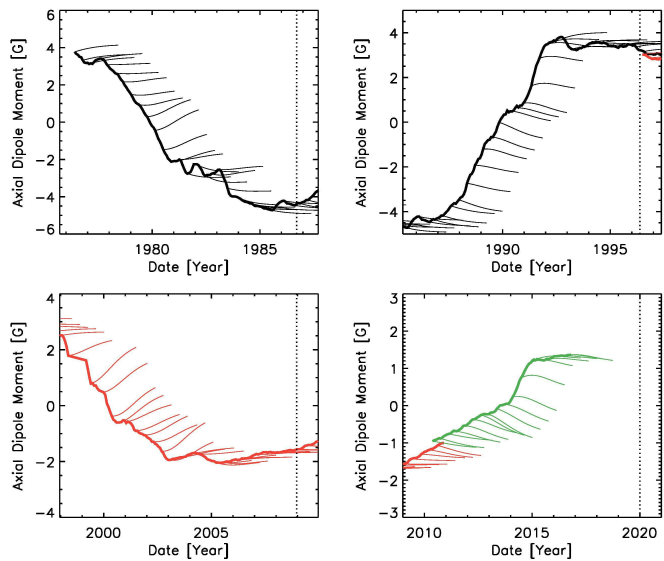}
 \caption{
 Time evolution of the observed
 and predicted axial dipole moment.
 The thick lines indicate the observed axial dipole moment
 (black: WSO, red: MDI, green: HMI).
 The thin lines indicate the prediction
 by the SFT without introducing new emerging flux
 starting from the observed synoptic magnetogram.
 The dotted lines indicate the timing of the cycle minimum.
 The minimum in Cycle 24 is assumed to be in 2020.
 \label{pred}
 }
\end{figure}

The thick solid lines in Figure \ref{pred} shows
the time evolution of the observed axial dipole moment
in Cycle 21, 22, 23, and 24.
As a precursor of the next solar cycle amplitude,
we use the axial dipole moment
$D=3/(4\pi)\int_{4\pi}B_R\sin\lambda{d}\Omega$,
where $B_R$ is the radial component of the magnetic field
and $\lambda$ is the latitude.
We find that the last three cycles (21--23)
have the periods (or plateaus) in which
the time evolution of the axial dipole moment
becomes very small near the end of each cycle.
This plateau continues at least approximately three years
before each sunspot cycle minimum.
It is also notable that the plateau in Cycle 23
continues exceptionally long time (from $\sim2004$ to $\sim2009$),
which is known as the extended cycle minimum
\citep[e.g.,][]{2014ApJ...792..142U,2015ApJ...808L..28J}.
We explain the thin solid lines in Section \ref{sec:model}.

\begin{figure}[!t]
 \includegraphics[width=\hsize]{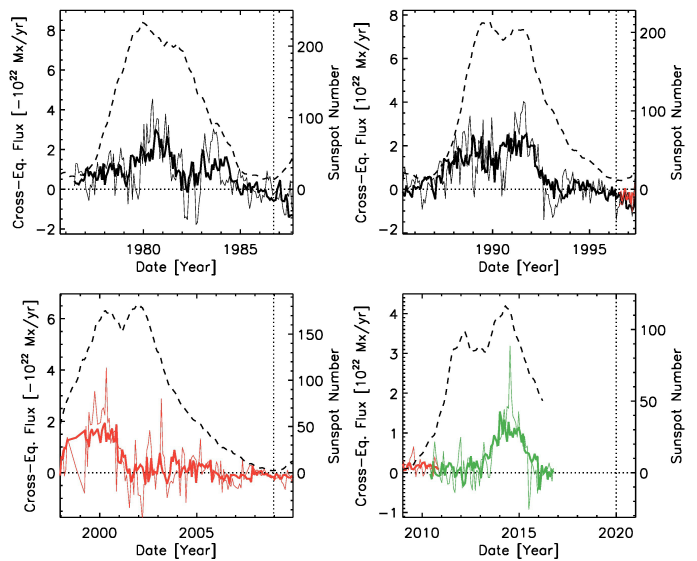}
 \caption{
 Time variation of the 13 month smoothed
 monthly total sunspot number (dashed lines)
 and the cross-equatorial flux transport (solid lines)
 in each sunspot cycle.
 The definition of the cross-equatorial flux transport
 is described in the body text.
 The cross-equatorial flux transport is smoothed
 by taking the moving average
 of 1 year (thin) and 2 years (thick).
 The dotted lines indicate the timing of the cycle minimum.
 The minimum in Cycle 24 is assumed to be in 2020.
 \label{ssn}
 }
\end{figure}

The cross-equatorial transport of the net magnetic flux
plays a crucial role for the build-up
of the axial dipole moment in the solar minimum
\citep{2013A&A...557A.141C}.
The temporal change of the axial dipole moment
is affected by the poleward transport of the active region
and the cross-equatorial transport of the magnetic flux.
On the other hand, the permanent change of
the axial dipole moment in the cycle minimum,
in which very small amount of the magnetic flux
remains in the low latitude,
is only produced by the cross-equatorial flux transport.

We define the cross-equatorial flux transport
as the time derivative of the net magnetic flux.
The net magnetic flux in the northern hemisphere
is defined as
\begin{equation}
 \Phi^N=R_\odot^2\int_{0}^{\pi/2}\cos{\lambda}d\lambda
 \int_{-\pi}^{\pi}d\phi B_R(\lambda,\phi),
\end{equation}
where $\phi$ is the longitude.
Because the magnetic monopole is not allowed,
the net flux in the northern hemisphere $\Phi^N$
exactly balances with the net flux in the southern hemisphere.
Thus, no hemispheric asymmetry exists in the net hemispheric flux.
The cross-equatorial transport of the net magnetic flux per unit time
is computed by the time derivative of the net flux $d\Phi^N/dt$.

Figure \ref{ssn} shows the comparison between
the cross-equatorial flux transport (solid lines)
and the monthly total sunspot number (dashed lines).
We find the significant number of sunspots
(several tens of percents of the maximum)
emerging during the plateau of the axial dipole moment
(approximately 3 years before each minimum; Figure \ref{pred}).
However, the cross-equatorial flux transport
is nearly zero during the axial dipole moment plateau
(typically 3 years before each minimum; Figure \ref{pred}).
This suggests that the newly emerged sunspots
does not contribute to the variation of the axial dipole moment
during the period.
This is surprising because the sunspots
appears at the low latitude
in the latter phase of the sunspot cycle,
which makes the cross-equatorial flux transport easily.
\tred{
We note that the cross-equatorial flux
based on the Kitt Peak Vacuum Telescope
of the US National Solar Observatory (NSO)
has larger time variation before the mid-1990s
\citep[see Figure 5 in ][]{2013A&A...557A.141C}
than that in the WSO data,
which makes it difficult to identify the existence
of the axial dipole moment plateau in the NSO data
in the end of Cycle 21.
The further analysis should be undertaken on this difference.
}
In the next section,
we further show that the time evolution
of the observed axial dipole moment near the end of each cycle
is well modeled by the SFT model without introducing
the new flux emergence.

\section{Modeling and Prediction\label{sec:model}}

Following the observational results in Section \ref{sec:observation},
we try to model the time evolution of the axial dipole moment
by the SFT model without introducing the new flux emergence.
We model the time evolution of the photospheric radial magnetic field
based on the Surface Flux Transport (SFT) model
\citep[e.g.,][]{1983IAUS..102..273S,1984SoPh...92....1D,
1989Sci...245..712W,1998ApJ...501..866V}.
The basic equation of the SFT code,
i.e., the two-dimensional advection-diffusion equation,
is azimuthally averaged to get the one-dimesional SFT equation
\citep[e.g.,][]{2007ApJ...659..801C}.
The equation is written as
\begin{equation}
 \frac{\partial B_R}{\partial t}
 +\frac{1}{R_\sun\sin\theta}
 \frac{\partial}{\partial\theta}
 \left(B_RV_\theta\sin\theta\right)
 =\frac{1}{R_\sun^2\sin\theta}
 \frac{\partial}{\partial\theta}
 \left(\eta\sin\theta\frac{\partial B_R}{\partial\theta}\right),
 \label{eq:basic}
\end{equation}
where $\theta$ is the colatitude.
The meridional flow $V_\theta$ is taken from \cite{1998ApJ...501..866V}.
The turbulent magnetic diffusivity $\eta$ is
assumed to be $250$ km$^2$/s \citep[e.g.,][]{2016ApJ...823L..22C}.
This one-dimensional SFT equation can describe
the evolution of the azimuthally averaged magnetic field
that is analytically identical to the original two-dimensional SFT equation
with the longitudinally and temporally constant
turbulent diffusivity, meridional flow, and differential rotation.
Because we only focus on
the azimuthally averaged axial dipole moment
or the axial magnetic dipole moment,
this one-dimensional SFT equation is sufficient for this study.
The equation is solved by the second-order
central finite difference scheme in space
and the second-order Strong Stability Preserving Runge-Kutta method in time.
The advection term is stabilized only
by the turbulent magnetic diffusion
and no additional artificial diffusion is used.
The latitudinal grid size is $5.5$ Mm in this study.

The thin solid lines in Figure \ref{pred} shows
the simulations of our SFT model (Eq. \ref{eq:basic})
that does not include the contribution
of the new flux emergence.
Each SFT simulation is started from each snapshot
of the longitudinally averaged synoptic magnetogram
and integrated for five years.
Near the cycle maximum,
the observed axial dipole moment greatly changes in time
and changes the sign of the axial dipole moment.
On the other hand, the axial dipole moment predicted by the SFT
evolves \tred{toward} to the value in the preceding minimum.
This behavior is caused because
the independent tilted active region
contributes to the global axial dipole moment
\citep[e.g.,][]{1991ApJ...375..761W,
2014ApJ...791....5J,2015SoPh..290.3189Y}.
When the latitude of the emergence is high enough,
this contribution is transient and eliminated
by the poleward transport of the active region.
When the active region emerged at low latitude
(as in the latter part of the solar cycle),
the magnetic flux of the preceding polarity
can be transported across the equator.
This contribution is not canceled by the poleward transport.
This feature is shown more clearly
in the HMI/MDI data than the WSO data
in which small active regions are flattened (or averaged)
by the lower spatial resolution.
During the plateau of the axial dipole moment near the end of each Cycle,
the time variation of the predicted dipole moment also becomes small.
The simulated axial dipole moment shows good agreement
with the observed dipole moment.
The small time variation
of the axial dipole moment during the plateau indicates
small cross-equatorial transport of the magnetic flux.
Although we do not show the WSO data in Cycle 23 and 24
for the visibility of Figure \ref{pred},
we note that the observed and predicted axial dipole moments
of the WSO data also exhibit the characteristics
similar to the MDI/HMI data.

\begin{figure}[!t]
 \includegraphics[width=\hsize]{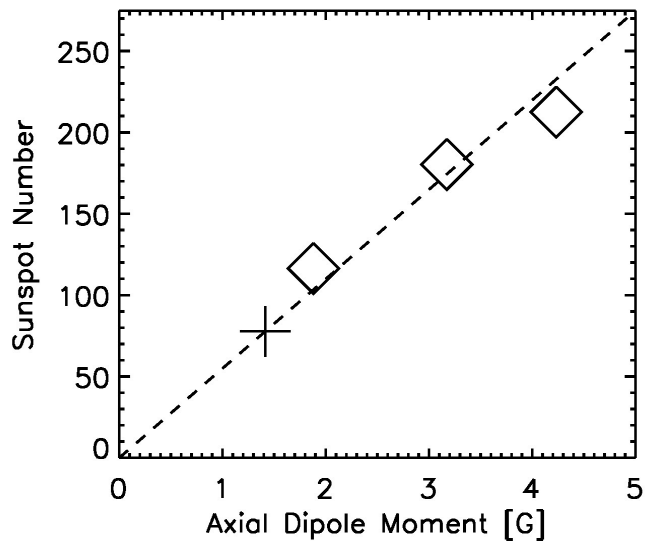}
 \caption{
 Maximum value of the 13 month smoothed
 monthly total sunspot number in each sunspot cycle
 (Cycle 22, 23, and 24; diamond)
 and the predicted cycle amplitude in Cycle 25 (cross)
 as a function of the axial dipole moment
 at the previous minimum
 predicted from the magnetogram
 observed three-year before the minimum.
 The least-square fit for Cycle 22, 23, and 24
 that across the point of origin
 is shown as the dashed line.
 The correlation coefficient
 for Cycle 22, 23, and 24 is 0.99.
 \label{cycle}
 }
\end{figure}

Under the assumption that the contribution of the emerging flux
is negligibly small in the period of several years
before each cycle minimum,
we predict the future axial dipole moment
and the cycle amplitude with the SFT model
without the new flux emergence.
As shown in Figure \ref{ssn},
the cross-equatorial flux transport in Cycle 24
becomes approximately zero from the beginning of 2016.
This result allows us to predict the axial dipole moment
in Cycle 24/25 minimum from the current observation.
Figure \ref{cycle} shows
the amplitude of the solar sunspot cycle
as a function of the three-year prediction
of the axial dipole moment in the cycle minimum.
The amplitude of each cycle
is measured by the 13 month smoothed
monthly total sunspot number.
The total sunspot number is averaged
over the six months before and after the cycle maximum.
The predicted axial dipole moment in the cycle minimum
is simulated from the synoptic magnetogram
3 years before the solar minimum
by the SFT neglecting new emergence of magnetic flux.
The axial dipole moment is averaged
over the six months before and after
the timing of 3 years before the solar minimum.
In Cycle 23/24 minimum (at the end of Cycle 23)
and Cycle 24/25 minimum,
multiple data sources (WSO, MDI, and HMI) are available.
In such cases, we use the average value of
the axial dipole moments independently
predicted from the available data sources.
The cycle maximum/minimum is defined
as the date in which the 13 month smoothed
monthly total sunspot number becomes maximum/minimum.

The predicted axial dipole moment
is highly correlated to the amplitude of the next cycle
with the correlation coefficient of 0.99
for Cycle 22, 23, and 24.
This proves the predictability
of the future solar cycle using our method.
We assume the proportional relation
between the predicted axial dipole moment of the minimum
and the maximum sunspot number in the next cycle
and apply the least-square fit for Cycle 22, 23, and 24.
Based on the relation,
we predict that the maximum total sunspot number in Cycle 25
will be 60--80 percent of Cycle 24.
The error range of the predicted value
comes from the averaging procedure to measure
the total sunspot number and the axial dipole moment,
the deviation from the proportional relation between them,
and the difference among the multiple instruments.
We also note the deviation of the observed axial dipole moment
exists between the WSO ($\sim 1.77$ G)
and HMI ($\sim 1.39$ G; bottom right panel in Figure \ref{pred})
data in 2015--2017 as a source of the prediction error.

\section{Discussion\label{sec:discussion}}

In this study, we have shown that the solar cycles 21--24
has the common characteristic
that the axial dipole moment hardly changes
during the period near the end of each cycle,
which we call the axial dipole moment plateau.
The cross-equatorial flux transport
becomes very small during the period.
This is also confirmed by showing that
the time evolution of the axial dipole moment
is well described by the simplified SFT model
without the new flux emergence.
The axial dipole moment predicted by the SFT model shows
high correlation to the amplitude in the next cycle,
which allows us to predict the amplitude of the Cycle 25.

We get a high correlation between the axial dipole moment
predicted three years before the cycle minimum
and the amplitude of the next cycle maximum.
We note that the similar high correlation
is achieved between the axial dipole moment
three years before the cycle minimum
and the amplitude of the next solar cycle.
This high correlation comes from the existence
of the plateau near the end of each solar cycle.
The result indicates the importance of the plateau
for the prediction of the future axial dipole moment
and the solar activity.
\cite{2013PhRvL.111d1106M} reported that
the cycle prediction based on the polar field measurement
perform well up to two years before minimum,
after which the success rate drops dramatically.
We further emphasize this point and show that,
although the significant number of sunspots are still observed,
the new emergence of the magnetic flux
does not contributes to the polar field near the end of the cycle.

We predict that the strength of the axial dipole moment at Cycle 24/25 minimum
will be several tens of percent weaker than the previous minimum.
This value is comparable but weakest among the predictions
based on the other surface flux transport models
\citep{2014ApJ...780....5U,2016ApJ...823L..22C,2016arXiv161105106H}.
It is natural that we get the weakest predicted value
because we neglect the contribution
of the new emergence of the magnetic flux in the simulation.

We find the existence of the plateau of the axial dipole moment
before the cycle minimum for last three cycles.
The duration of the plateau ranges
from $\sim3$ years in Cycle 21 to $\sim6$ years in Cycle 23.
Although all of the cycles investigated in this study
have the plateau lasting several years,
we can not deny the possibility
that the last three cycles were the special cases
in the long history of the sunspot cycles.
\cite{2012ApJ...753..146M} reported the polar magnetic field
estimated from the polar faculae measurements during 100 years.
Their estimate of the polar field suggests
that not all of the cycles show the plateau reported in our study
in the past 100 years.
Because the difference between the temporal evolution
of the faculae count and the axial dipole moment is significant,
we need further studies to clarify the universality of the plateau.

The physical origin of the axial dipole moment plateau
is not clear at present.
Because the significant amount of the sunspot appears
during the plateau, we need explanations for
such small cross-equatorial flux transport during the period.
The anomalously weak polar field
and the extended minimum at the end of Cycle 23
has been studies by various authors
\citep[see also a review by][]{2015LRSP...12....5P}.
One explanation of the small polar field
at Cycle 23/24 minimum
is that the emergence of the active region
with the small tilt angle or the opposite polarity
causes the source of the small cross-equatorial flux transport
\citep{2015ApJ...808L..28J}.
The other explanation is the variation of the meridional flow.
The converging motion of the meridional flow
in the activity band or the active region inflow
\citep[e.g.,][]{2002ApJ...570..855H,2004ApJ...603..776Z,2010Sci...327.1350H}
in the low latitude will prevent the cross-equatorial transport
of the magnetic flux
\citep[e.g.,][]{2006ESASP.624E..12D,2012A&A...548A..57C,2017A&A...597A..21M}.
The high gradient of the meridional flow near the equator
is also the candidate of the small cross-equatorial flux transport
\citep{2008SoPh..252...19S,2009ApJ...707.1372W,2013SSRv..176..289J}.
Although the meridional flow variation
on the Cycle 23/24 minimum is not suitable
to explain the observed axial dipole moment
\citep{2014ApJ...792..142U},
the effect on the flux transport
should be considered to explain the plateau.

The deviation from the SFT model is also
a candidate of the origin of the axial dipole moment plateau.
In the SFT model, the only effect of meridional flow is
the horizontal (poleward) transport of the surface magnetic field.
However, some studies indicated the possible role of
the vertical (radial) component of the meridional circulation
for the evolution of the surface magnetic field
\citep[e.g.,][]{1994A&A...291..975D,2017ApJ...835...39H}.
On the other hand, \cite{2012A&A...542A.127C}
suggested that the effect of the vertical flow
of the meridional circulation is suppressed when
the sufficiently strong downward magnetic pumping
exists near the solar surface.
Although the actual strength of the magnetic pumping
near the solar surface is not well known,
\cite{2003IAUS..210..169S} suggested that
the convective motion near the surface
will expel the magnetic flux rapidly
under the strong density stratification,
which implies the strong magnetic pumping.
Further studies is required for the quantitative evaluation
of the effect of the multi-dimensional meridional flow
to the evolution of the surface magnetic field.

\begin{acknowledgements}
We would like to thank the anonymous referee for helpful suggestions.
Wilcox Solar Observatory data used in this study was obtained via the
web site http://wso.stanford.edu.
The Wilcox Solar Observatory is currently supported by NASA,
SOHO is a project of international cooperation between ESA and NASA.
The SDO/HMI data used are courtesy of NASA and the SDO/HMI team.
The sunspot records are courtesy of
WDC-SILSO, Royal Observatory of Belgium, Brussels.
This work is supported by MEXT/JSPS KAKENHI Grant Number 15H05816.
HH is supported by MEXT/JSPS KAKENHI Grant Number
JP16K17655, JP16H01169.
This work is carried out by the joint research program of the Institute for
Space-Earth Environment Research (ISEE), Nagoya University.
\end{acknowledgements}


\end{document}